\documentclass[12pt]{article}
\usepackage{times,epsfig,amssymb}
\voffset     = - 0.10 in
\begin{document}

\begin{center}
{\large\bf
Hadron Energy Reconstruction for the ATLAS Calorimetry
in the Framework of the Non-parametrical Method
}


\smallskip

ATLAS Collaboration (Calorimetry and Data Acquisition)
\end{center}

{\small

  \noindent
S.~Akhmadaliev$^{21}$, P.~Amaral$^{15,a}$, G.~Ambrosini$^{9}$,
A.~Amorim$^{15,a}$, K.~Anderson$^{10}$, M.L.~Andrieux$^{14}$,
B.~Aubert$^{1}$, E.~Aug\'{e}$^{22}$, F.~Badaud$^{11}$,
L.~Baisin$^{9}$
% ...
{\raggedright

}

\bigskip

\begin{center}
$^{1}$     LAPP, Annecy, France \\
% ...
$^{9}$   CERN, Geneva, Switzerland   \\
$^{10}$  University of Chicago, USA  \\
$^{11}$  LPC Clermont--Ferrand, Universit\'e Blaise Pascal / CNRS--IN2P3,
         France  \\
% ...
$^{14}$  ISN, Universit\'e Joseph Fourier /CNRS-IN2P3, Grenoble, France   \\
$^{15}$  (a) LIP-Lisbon and  FCUL-Univ. of Lisbon,
         (b) LIP-Lisbon and  FCTUC-Univ. of Coimbra, \\
         (c) LIP-Lisbon and Univ. Catolica Figueira da Foz \\
% ...
$^{21}$  Budker Institute of Nuclear Physics, Novosibirsk, Russia \\
$^{22}$  LAL, Orsay, France \\
% ...
\end{center}

}


\newpage
\bigskip
\bigskip

\begin{abstract}
This paper discusses hadron energy reconstruction for the ATLAS
barrel prototype combined calorimeter (consisting of a
lead-liquid argon electromagnetic part and an iron-scintillator
hadronic part) in the framework of the non-pa\-ra\-met\-ri\-cal
method. The non-parametrical me\-thod utilizes only the known
$e/h$ ratios and the electron calibration constants and does not
require the determination of any parameters by a minimization
technique. Thus, this technique lends itself to an easy use in a
first level trigger. The reconstructed mean values of the hadron
energies are within $\pm 1\%$ of the true values and the
fractional energy resolution is $[(58\pm3)\%
/\sqrt{E}+(2.5\pm0.3)\%]\oplus (1.7\pm0.2)/E$. The value of the
$e/h$ ratio obtained for the electromagnetic compartment of the
combined calorimeter is $1.74\pm0.04$ and agrees with the
prediction that $e/h > 1.7$ for this electromagnetic calorimeter.
Results of a study of the longitudinal hadronic shower development
are also presented. The data have been taken in the H8 beam line
of the CERN SPS using pions of energies from  10 to 300 GeV.
\vskip 5mm \noindent {\bf Codes PACS:} 29.40.Vj, 29.40.Mc,
29.85.+c.\\
{\bf Keywords:}
Calorimetry,
Combined Calorimeter,
Shower Counter,
Compensation,
Energy Measurement,
Computer Data Analysis.
\end{abstract}

\newpage
\section{Introduction}

\begin{figure}[tbph]
\begin{center}
\mbox{\epsfig{figure=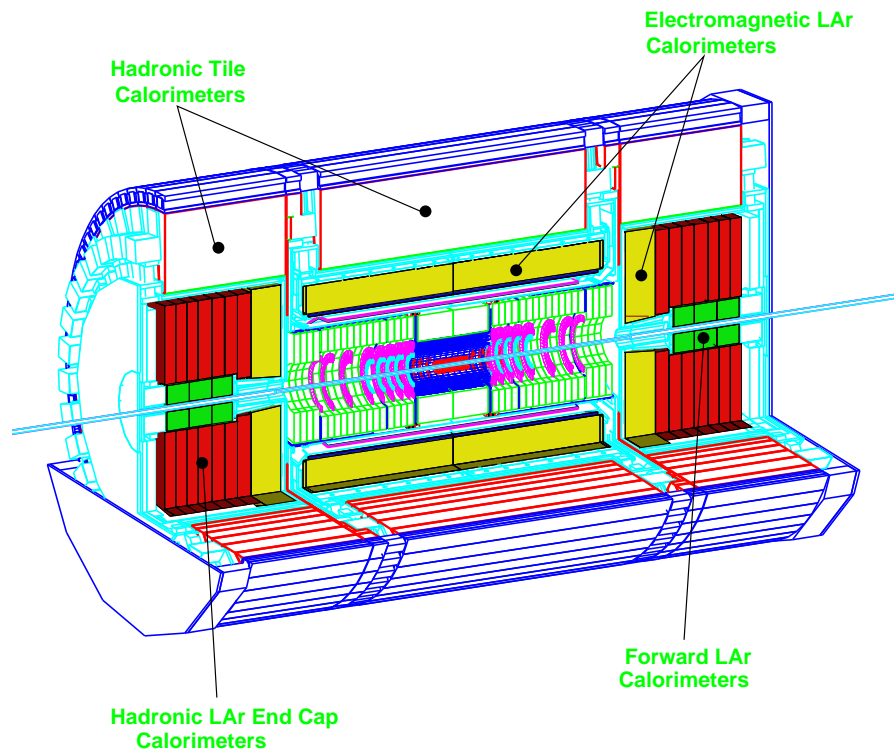,width=0.95\textwidth}}
 \end{center}
 \caption{
        Three-dimensional cutaway view of the ATLAS calorimeters.
        }
\label{f-00}
\end{figure}

The key question for calorimetry in general, and hadronic calorimetry in
particular, is that of energy reconstruction.
This question becomes especially important when a hadronic calorimeter
has a complex structure incorporating electromagnetic and hadronic
compartments with different technologies.
This is the case for the central (barrel) calorimetry of the
ATLAS detector which has the electromagnetic liquid argon accordion and hadronic
iron-scintillator Tile calorimeters
\cite{atcol94,TILECAL96,LARG96}.
A view of the ATLAS detector, including the two calorimeters, is shown
in Fig.\ \ref{f-00}.

In this paper, we describe a non-parametrical method of  energy
reconstruction for a combined calorimeter known as the $e/h$ method,
and demonstrate its performance using the test beam data from the
ATLAS combined prototype calorimeter.
For the energy reconstruction and description of the longitudinal
development of a hadronic shower, it is necessary to know the $e/h$
ratios, the degree of non-compensation, of these  calorimeters.
Detailed information about the $e/h$ ratio for the ATLAS Tile barrel
calorimeter is  presented in
\cite{TILECAL96,ariztizabal94,juste95,budagov96-72,kulchitsky99-12}
while much less was done so far for the liquid argon electromagnetic
calorimeter \cite{stipcevic94,stipcevic93,Gingrich95}.
An additional aim of the present work, then, is to also determine
the value of the $e/h$ ratio for the electromagnetic compartment.

Another important question for hadron calorimetry is that
relating to the longitudinal development of hadronic showers.
This question is especially important for a combined calorimeter
because of the different degrees of non-compensation for the
separate calorimeter compartments. Information about the
longitudinal hadronic shower development is very important for
fast and full hadronic shower simulations and for fast energy
reconstruction in a first level trigger. This work is also
devoted to the study of the longitudinal hadronic shower
development in the ATLAS combined calorimeter.

This work has been performed using the 1996 combined test beam
data \cite{comb96,cobal98} taken in the H8 beam line of the CERN SPS
using pions of  energies from 10 to 300 GeV.

\section{Combined Calorimeter}

\begin{figure}[tbph]
\begin{center}
\mbox{\epsfig{figure=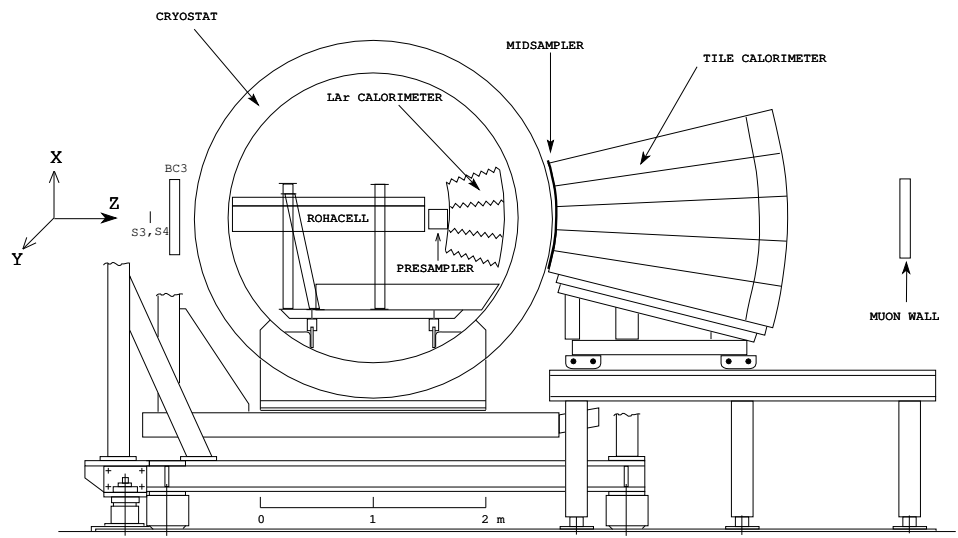,width=0.95\textwidth
}}
\end{center}
 \caption{
        Schematic layout of the experimental setup for the
        combined
        LAr and Tile calorimeters run (side view).
        The S3 and S4 are scintillation counters,
        the BC3 is a beam proportional chamber, and
        the midsampler and the ``muon wall'' are scintillation hodoscopes.
        }
\label{fv1}
\end{figure}

The combined calorimeter prototype setup is shown in Fig.\ \ref{fv1},
along with a definition of the coordinate system used for the test beam.
The LAr calorimeter prototype is housed inside a
cryostat with the hadronic Tile calorimeter prototype located downstream.

The beam line is in the $YZ$ plane at 12 degrees from the $Z$
axis. With this angle the two calorimeters have an active
thickness of 10.3 interaction lengths ($\lambda_I$). The beam
quality and geometry were monitored with a set of scintillation
counters S1 -- S4, beam wire chambers BC1 -- BC3 and trigger
hodoscopes (midsampler) placed downstream of the cryostat. To
detect punchthrough particles and to measure the effect of
longitudinal leakage a ``muon wall'' consisting of 10
scintillator counters (each 2 cm thick) was located behind the
calorimeters at a distance of about 1 metre.

The liquid argon electromagnetic calorimeter prototype consists
of a stack of three azimuthal modules, each module spanning
$9^\circ$ in azimuth and extending over 2000 mm along the $Y$
direction. The calorimeter structure is defined by 2.2 mm thick
steel-plated lead absorbers folded into an accordion shape and
separated by 3.8 mm gaps filled with liquid argon. The signals
are collected by three-layer copper-polyamide electrodes located
in the gaps. The calorimeter extends from an inner radius of 1315
mm to an outer radius of 1826 mm, representing (in the $Z$
direction) a total of 25 radiation lengths ($X_0$), or 1.22
$\lambda_I$ for protons. The calorimeter is longitudinally
segmented into three compartments of $9\ X_0$, $9\ X_0$ and $7\
X_0$, respectively. The $\eta\times\phi$ segmentation is
$0.018\times 0.02$ for the first two longitudinal compartments
and $0.036\times 0.02$ for the last compartment. Each read-out
cell has full projective geometry in $\eta$ and $\phi$. The
cryostat has a cylindrical shape, with a 2000 mm internal
diameter (filled with liquid argon), and consists of an 8 mm
thick inner stainless-steel vessel, isolated by 300 mm of
low-density foam (Rohacell), which is itself covered by a 1.2 mm
thick aluminum outer wall. A presampler was mounted in front of
the electromagnetic calorimeter. The presampler has fine strips in
the $\eta$ direction and covers $\thickapprox 11\times 8$ in
$\eta\times\phi$ LAr calorimeter cells in the region of the beam
impact. The active depth of liquid argon in the presampler was 10
mm and the strip spacing 3.9 mm. Early showers in the liquid
argon were kept to a minimum by placing light foam material
(Rohacell) in the cryostat upstream of the LAr electromagnetic
calorimeter. The total amount of material between BC3 and LAr
calorimeter is near $0.2 \lambda_I$.  More details about this
prototype can be found in \cite{atcol94,Gingrich95}.

\begin{figure*}[tbph]
     \begin{center}
        \begin{tabular}{|c|}
        \hline
\mbox{\epsfig{figure=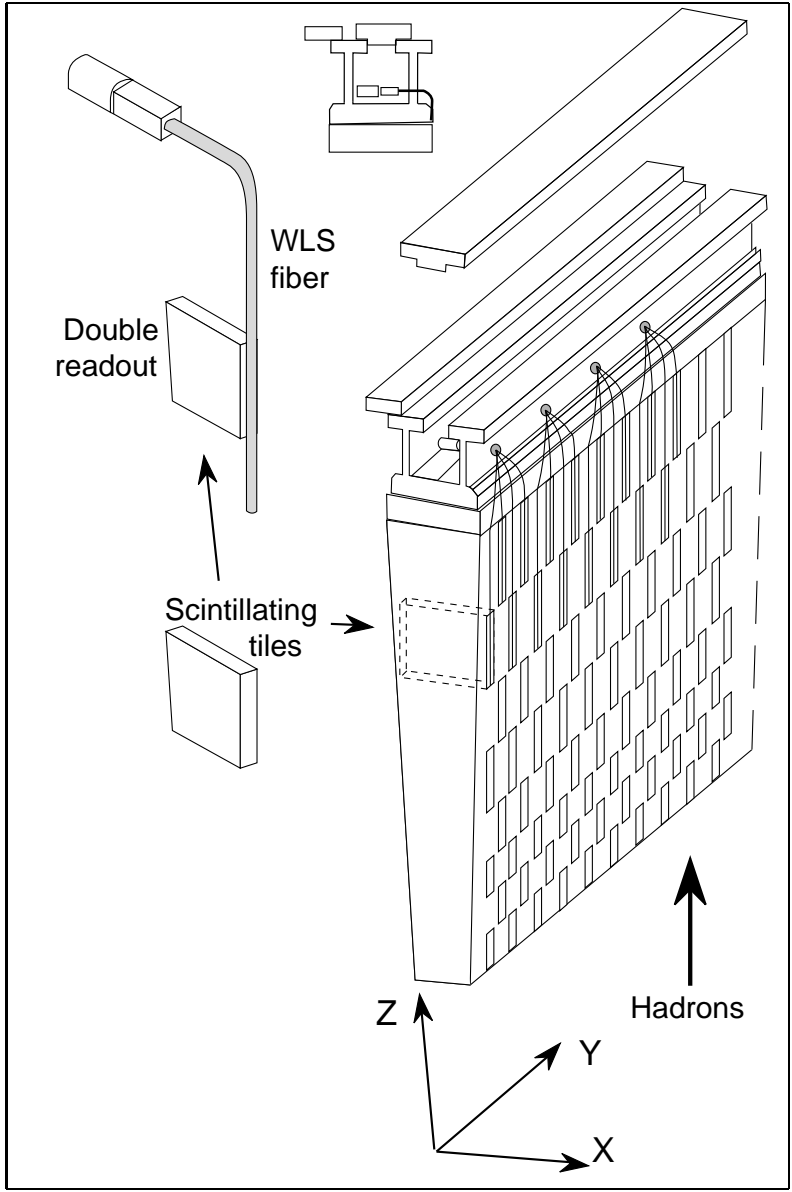,width=0.75\textwidth
}}
\\
        \hline
        \end{tabular}
     \end{center}
       \caption{
       Conceptual design of a Tile calorimeter module.
       \label{fig:tc}}
\end{figure*}

The hadronic Tile calorimeter is a sampling device which uses
steel as the absorber and scintillating tiles as the active
material \cite{TILECAL96}. A conceptual design of this
calorimeter geometry is shown in Fig.\  \ref{fig:tc}. The
innovative feature of the design is the orientation of the tiles
which are placed in planes perpendicular to the $Y$ direction
\cite{gild91}. The absorber structure is a laminate of steel
plates of various dimensions stacked along $Y$. The basic
geometrical element of the stack is denoted as a period. A period
consists of a set of two master plates (large trapezoidal steel
plates, 5 mm thick, spanning along the entire $Z$ dimension) and
one set of spacer plates (small trapezoidal steel plates, 4 mm
thick, 100 mm wide along $Z$). During construction, the
half-period elements are pre-assembled starting from an
individual master plate and the corresponding 9 spacer plates.
The relative position of the spacer plates in the two half
periods is staggered in the $Z$ direction, to provide pockets in
the structure for the subsequent insertion of the scintillating
tiles. Each stack, termed a  module, spans $2 \pi / 64$ in the
azimuthal angle ($X$ dimension), 1000 mm in the $Y$ direction and
1800 mm in the $Z$ direction (about 9 $\lambda_{I}$ or about 80
$X_0$). The module front face, exposed to the beam particles,
covers 1000$\times$200 mm$^2$. The scintillating tiles are made
out of polystyrene material of thickness 3 mm, doped with
scintillating and wavelength-shifting dyes. The iron to
scintillator ratio is $4.67 : 1$ by volume. The tile calorimeter
thickness along the beam direction at the incidence angle of
$12^{\circ}$ (the angle between the incident particle direction
and the normal to the calorimeter front face) corresponds to 1.5
m of iron equivalent length.

Wavelength shifting fibers collect the scintillation light from
the tiles at both of their open (azimuthal) edges and transport
it to photo-multipliers (PMTs) at the periphery of the calorimeter
(Fig.\ \ref{fig:tc}). Each PMT views a specific group of tiles
through the corresponding bundle of fibers. The prototype Tile
calorimeter used for this study is composed of five modules
stacked in the $X$ direction, as shown in Fig.\  \ref{fv1}.

The modules are longitudinally segmented (along $Z$) into four
depth segments. The readout cells have a lateral dimension of 200
mm along $Y$, and longitudinal dimensions of 300, 400, 500, 600
mm for depth segments 1 -- 4, corresponding to 1.5, 2, 2.5 and 3
$\lambda_{I}$, respectively. Along the $X$ direction, the cell
sizes vary between about 200 and 370 mm depending on the $Z$
coordinate (Fig.\ \ref{fv1}). More details of this prototype can
be found in
\cite{atcol94,berger95,bosman93,ariztizabal94,amaral00,budagov-97-127}.
The energy release in 100 different cells was recorded for each
event \cite{berger95}.

The data have been taken in the H8 beam line of the CERN SPS
using pions of energy 10, 20, 40, 50, 80, 100, 150 and 300 GeV.
We have applied some cuts similar to \cite{comb96,cobal98} in
order to eliminate the non-single track pion events, the beam
halo, the events with an interaction before the liquid argon
calorimeter, and the electron and muon events. The set of cuts
adopted is as follows: single-track pion events were selected by
requiring the pulse height of the beam scintillation counters and
the energy released in the presampler of the electromagnetic
calorimeter to be compatible with that for a single particle; the
beam halo events were removed with appropriate cuts on the
horizontal and vertical positions of the incoming track impact
point and the space angle with respect to the beam axis as
measured with the beam chambers; a cut on the total energy
rejects incoming muons.

\section{The $e/h$\  Method of Energy Reconstruction}

An hadronic shower in a calorimeter can be seen as an overlap of
a pure electromagnetic and a pure hadronic component. In this
case an incident hadron energy is $E = E_e + E_h$. The calorimeter
response, $R$, to these two components is usually different
\cite{wigmans88,groom89}  and can  be written as:
\begin{equation}
R =  e \cdot E_e + h \cdot E_h\ , \label{ev9}
\end{equation}
where $e$ ($h$) is a coefficient to rescale the electromagnetic
(hadronic) energy content to the calorimeter response. A fraction
of an electromagnetic energy of a hadronic shower is $f_{\pi^0} =
E_e/E$, than $R = e\cdot f_{\pi^0}\cdot E + h\cdot (E-
f_{\pi^0}\cdot E)= e\cdot [1+ (e/h -1)\cdot f_{\pi^0}]/(e/h) \cdot
E$. From this one can gets formulae for an incident energy
\begin{equation}
E = \frac{1}{e} \cdot \Biggl(\frac{e}{\pi}\Biggr)  \cdot R \ ,
\label{ev16}
\end{equation}
where
\begin{equation}
\Biggl(\frac{e}{\pi}\Biggr)=\frac{e/h}{1+(e/h-1)\cdot f_{\pi^0}}
\ . \label{ev10}
\end{equation}
The dependence of $f_{\pi^0}$ from the incident hadron energy can
be parameterized as in Ref.\ \cite{wigmans91}:
\begin{equation}
f_{\pi^0} = k \cdot \ln{E}\ .  \label{ev10-0}
\end{equation}

In the case of the combined setup described in this paper, the
total energy is reconstructed as the sum of the energy deposit in
the electromagnetic compartment ($E_{LAr}$),  the deposit in the
hadronic calorimeter ($E_{Tile}$),  and that in the passive
material between the LAr and Tile calorimeters ($E_{dm}$).
Expression (\ref{ev16}) can then be rewritten as:
\begin{equation}
E =  E_{LAr} + E_{dm} + E_{Tile}
=\frac{1}{e_{LAr}}\Biggl(\frac{e}{\pi}\Biggr)_{LAr} R_{LAr}
 + E_{dm}
 + \frac{1}{e_{Tile}}\Biggl(\frac{e}{\pi}\Biggr)_{Tile} R_{Tile}\ ,
\label{ev7}
\end{equation}
where $R_{LAr}$ ($R_{Tile}$) is the measured response of the LAr
(Tile) calorimeter compartment and $1/e_{Tile}$ and $1/e_{LAr}$
are energy calibration constants for the LAr and Tile
calorimeters respectively \cite{comb96}.

Similarly to the procedure in Refs.\ \cite{comb96,combined94}, the
$E_{dm}$ term, which accounts for the energy loss in the dead
material between the LAr and Tile calorimeters, is taken to be
proportional to the geometrical mean of the energy released in
the third depth of the electromagnetic compartment and the first
depth of the hadronic compartment ($E_{dm} = \alpha \cdot
\sqrt{E_{LAr, 3} \cdot E_{Tile, 1}}$). The validity of this
approximation has been tested using a Monte Carlo simulation
along with a  study of  the correlation between the energy
released in the midsampler and the $E_{dm}$
\cite{cobal98,bosman99,PTDR99}.

The  ratio $(e/h)_{Tile}=1.30\pm0.03$ has been  measured in a
stand-alone test beam run \cite{budagov96-72} and is used to
determine the $(e/\pi)_{Tile}$ term in equation \ref{ev7}. To
determine the value of the $1/e_{Tile}$ constant we selected
events which started showering only in the hadronic compartment,
requiring that the energy deposited in each sampling of the LAr
calorimeter and in the midsampler is compatible with that of a
single minimum ionization particle. The result is $1/e_{Tile}=
0.145\pm0.002$.

\begin{figure*}[tbph]
\begin{center}
\begin{tabular}{cc}
\mbox{\epsfig{figure=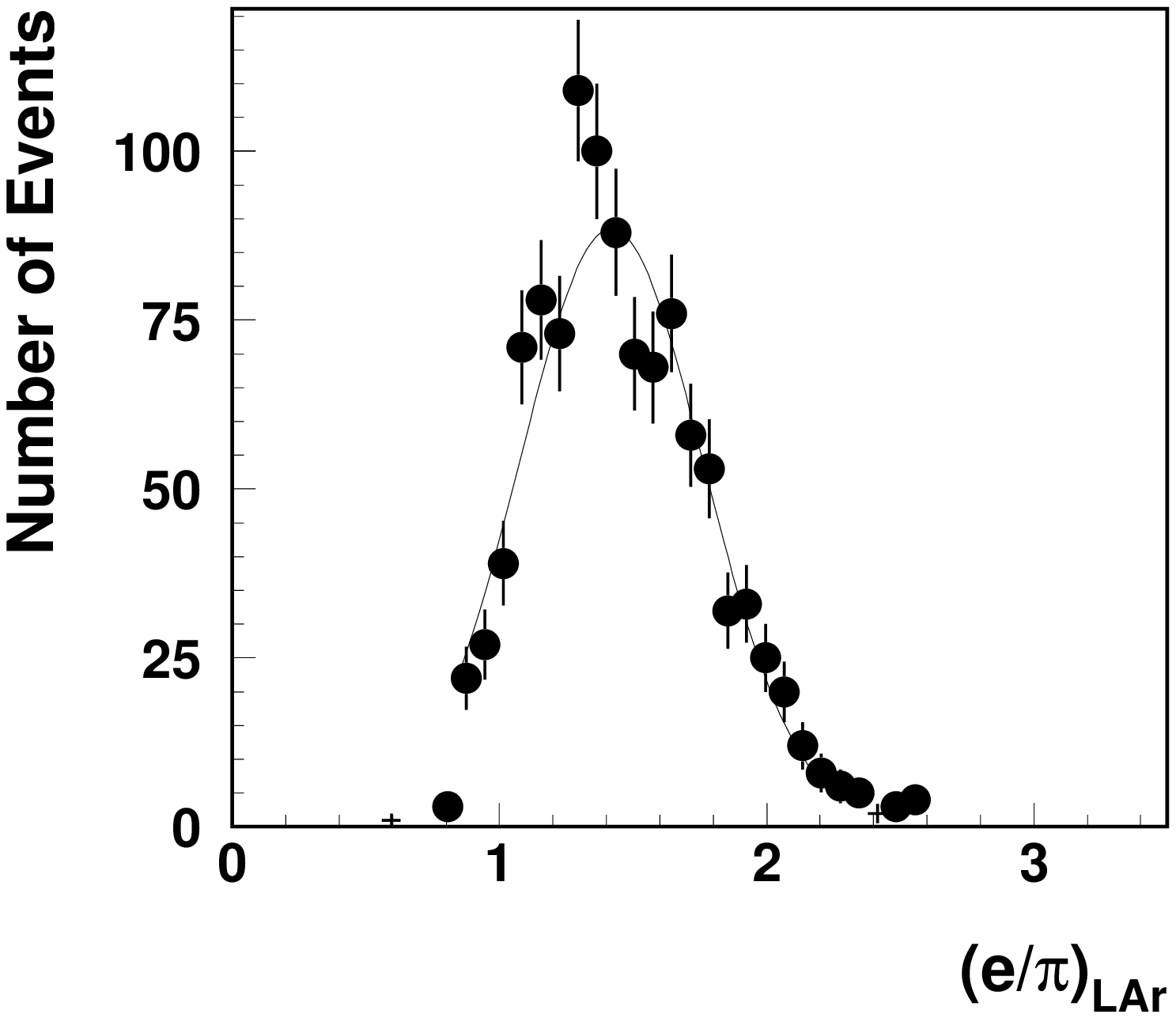,width=0.45\textwidth}}
&
\mbox{\epsfig{figure=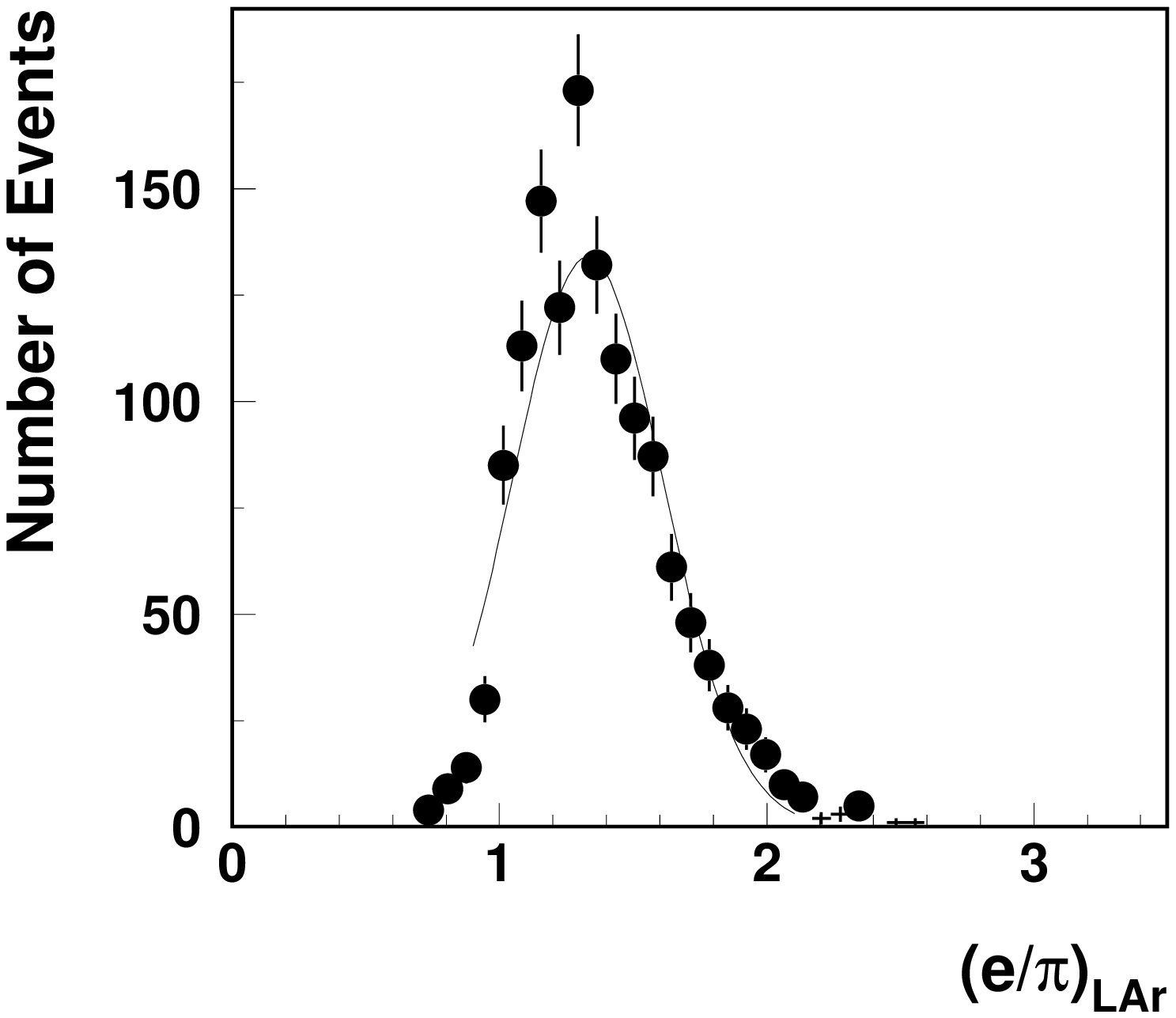,width=0.45\textwidth}}
\\
\mbox{\epsfig{figure=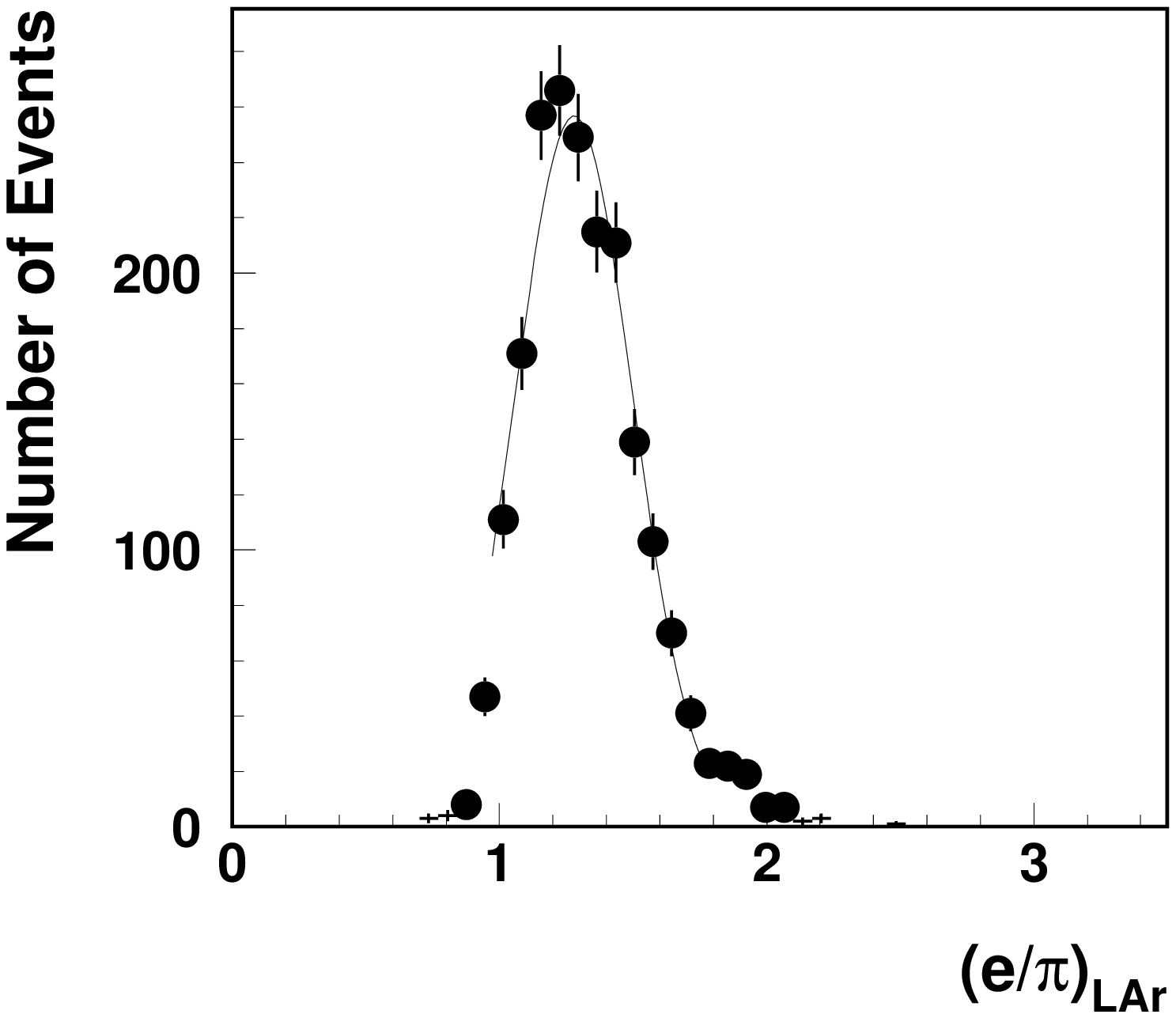,width=0.45\textwidth}}
&
\mbox{\epsfig{figure=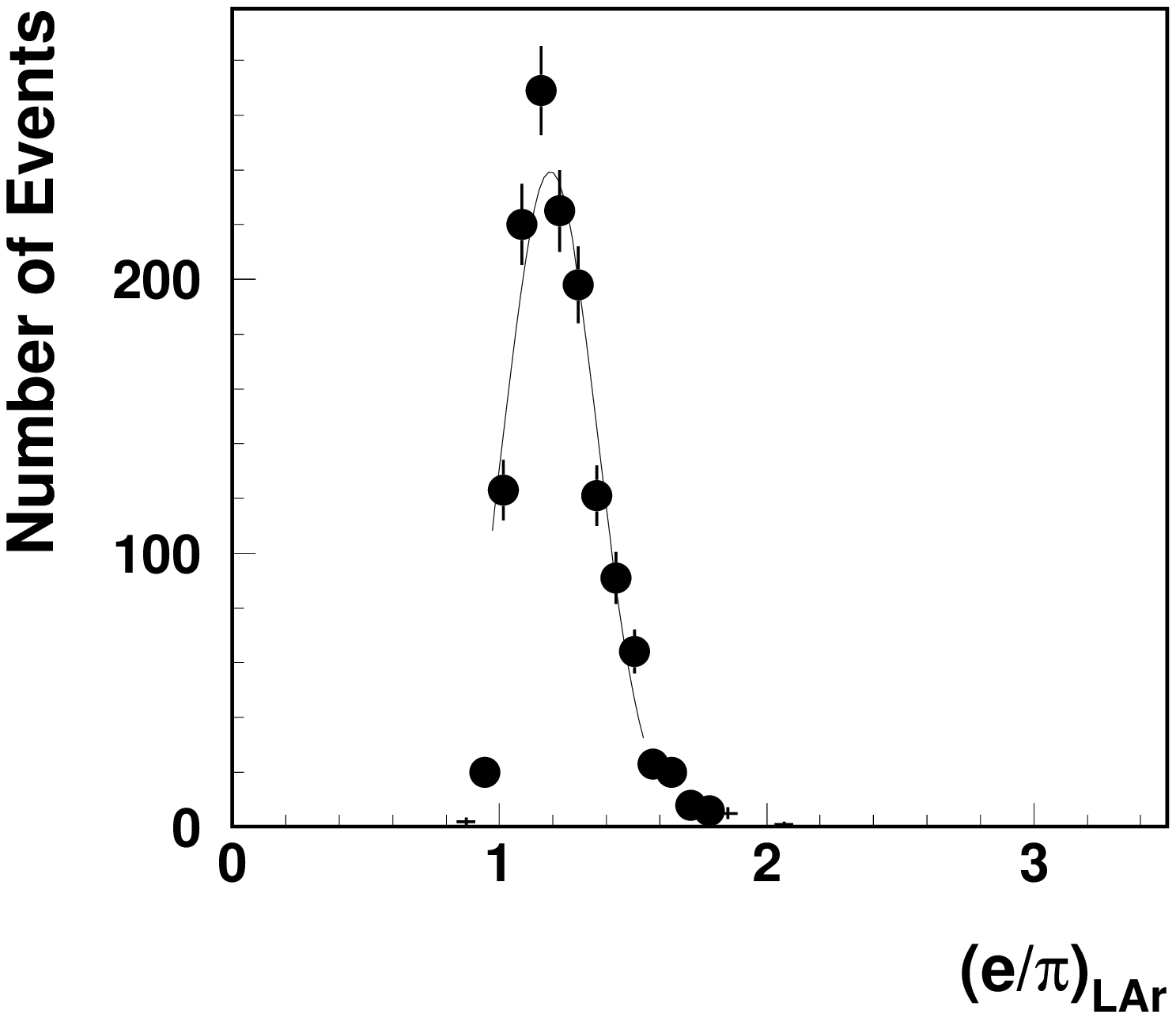,width=0.45\textwidth}}
\\
\end{tabular}
\end{center}
       \caption{The distributions of the $(e/\pi)_{LAr}$ ratio
                for beam energies of 20 and 50 GeV
                (top row, left to right),
                and beam energies of 100 and 300 GeV
                (bottom row, left to right).
       \label{fv2}}
\end{figure*}

\begin{figure*}[tbph]
\begin{center}
\begin{tabular}{c}
\mbox{\epsfig{figure=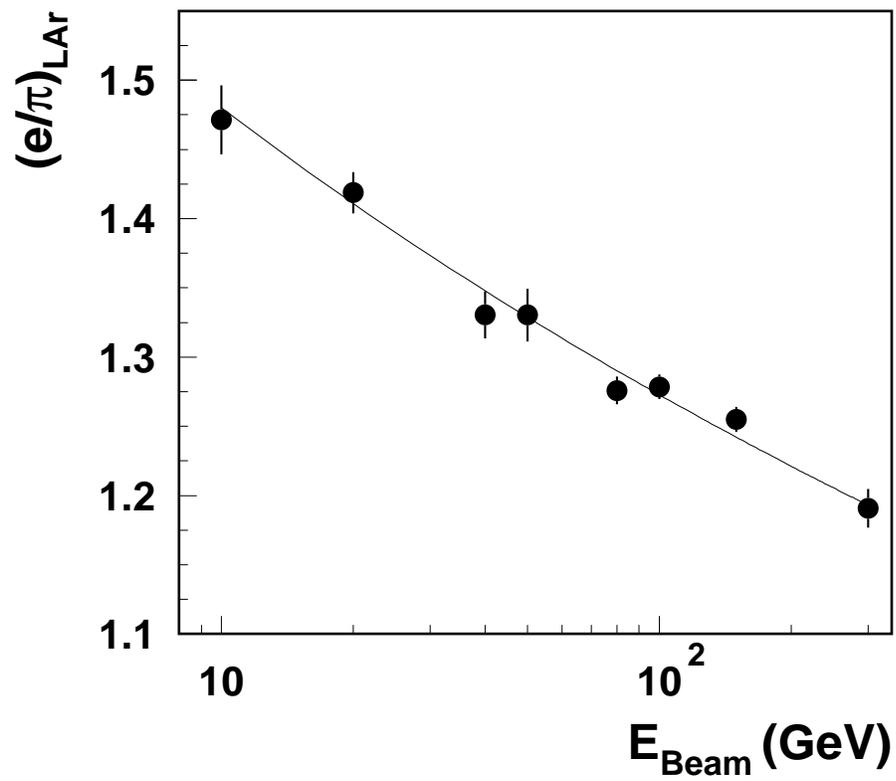,width=0.95\textwidth
}}
\\
\end{tabular}
\end{center}
\caption{The mean value of the $(e/\pi)_{LAr}$ ratio as a function
of the beam energy. The curve  is the result of a fit of equations
(\ref{ev10}) and (\ref{ev10-0}). \label{fv3}}
\end{figure*}

The response of the LAr calorimeter has already been calibrated
to the electromagnetic scale; thus the constant $1/e_{LAr} = 1$
\cite{comb96,cobal98}. The value of $(e/h)_{LAr}$ has been
evaluated using the data from this beam test, selecting   events
with  well developed hadronic showers in the electromagnetic
calorimeter, i.e. events with more than 10\% of the beam energy
in the electromagnetic ca\-lo\-ri\-me\-ter. Using the expression
(\ref{ev7}), the $(e/\pi)_{LAr}$ ratio can be written as:
\begin{equation}
        \Biggl( \frac{e}{\pi} \Biggr)_{LAr} =
        \frac{E_{beam} - E_{dm} - E_{Tile}}{R_{LAr}/e_{LAr}}\ .
\label{ev1}
\end{equation}
Fig.\  \ref{fv2} shows the distributions of the $(e/\pi)_{LAr}$
ratio for different energies, and the mean values of these
distributions are plotted in Fig.\ \ref{fv3} as a function of the
beam energy. From a fit to this distribution using expression
(\ref{ev10}) and (\ref{ev10-0}) we obtain
$(e/h)_{LAr}=1.74\pm0.04$ and $k = 0.108\pm0.004$, thereby taking
$(e/h)_{LAr}$ to be energy independent. For a fixed value of the
parameter $k = 0.11$ \cite{wigmans91}, the result is $(e/h)_{LAr}
= 1.77\pm0.02$. The quoted errors are the statistical ones
obtained from the fit. The systematic error on the $(e/h)_{LAr}$
ratio, which is a consequence of the uncertainties in the input
constants used in the equation (\ref{ev1}) as well as of the
shower development selection criteria, is estimated to be
$\pm0.04$.

\begin{figure}[tbh]
\begin{center}
\mbox{\epsfig{figure=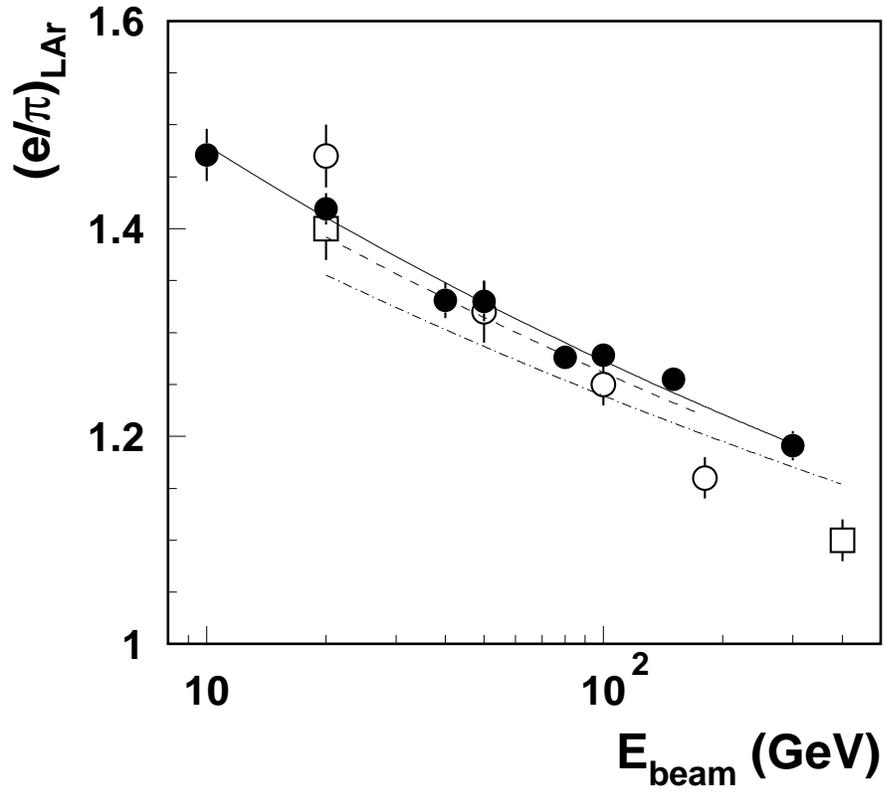,width=0.95\textwidth}}
\end{center}
\caption{The $(e/\pi)_{LAr}$ ratios as a function of the beam
energy. for $e/h$ method (black circles) and for weighting method
(open circles for \cite{stipcevic93} and open squares for
\cite{Gingrich95}). The lines are the result of a fit of equations
(\ref{ev10}) and (\ref{ev10-0}) with free $e/h$ parameter and
$k=0.11$: solid line is for our data, dashed line is for the
\cite{stipcevic93} data and dash-doted line is for the
\cite{Gingrich95} data. \label{fv3-00}}
\end{figure}

Figure \ref{fv3-00} compares our values of the $(e/\pi)_{LAr}$
ratio to the ones obtained in Refs.\
\cite{stipcevic94,stipcevic93,Gingrich95} using a weighting
method. The results are in good agreement below 100 GeV but
disagree above this energy because the weighting method leads to
a distortion of the $(e/\pi)_{LAr}$ ratios. Despite this
disagreement, fitting expression (\ref{ev10}) to the old data
leads to $(e/h)_{em} = 1.73\pm0.10$ for \cite{stipcevic93} and
$(e/h)_{em} = 1.64\pm0.18$ for \cite{Gingrich95} (parameter $k$
fixed at 0.11). These values are in agreement with our result
within error bars.

In the Ref.\ \cite{wigmans91} it was demonstrated that the $e/h$
ratio for non-uranium ca\-lo\-ri\-me\-ters with high-$Z$ absorber
material is satisfactorily described by the formula:
\begin{equation}
        \frac{e}{h}=\frac{e/mip}{0.41 + f_n \cdot n/mip}\ ,
\label{wig}
\end{equation}
where $f_n$ is a constant determined by the $Z$ of the absorber
(for lead $f_n=0.12$) \cite{wigmans87,wigmans98}, and $e/mip$ and
$n/mip$ represent the calorimeter response to electromagnetic
showers and to MeV-type neutrons, respectively. These responses
are normalized to the one for minimum ionizing particles. The
Monte Carlo calculated $e/mip$ and $n/mip$ values \cite{wigmans88}
for the lead liquid argon electromagnetic calorimeter
\cite{costa91} are $e/mip = 0.78$ and $n/mip < 0.5$, leading to
$e/h > 1.66$. The measured value of the $(e/h)_{em}$ ratio agrees
with this prediction. Using expression (\ref{wig}) and measured
value of $e/h$ we can find that $n/mip$ is $\simeq 0.3$.

Formula (\ref{wig}) indicates that $e/mip$ is very important for
understanding compensation in lead liquid argon calorimeters. The
degree of non-com\-pen\-sa\-ti\-on increases when the sampling
frequency is also increased \cite{wigmans87}. A large fraction of
the electromagnetic energy is deposited through very soft
electrons ($E < 1$ MeV) produced by Compton scattering or the
photoelectric effect. The cross sections for these processes
strongly depend on $Z$ and practically all these photon
conversions occur in the absorber material. The range of the
electrons produced in these processes is very short, $\sim 0.7$
mm for 1 MeV electron in lead. Such electrons only contribute to
the calorimeter signal if they are produced near the boundary
between the lead and the active material. If the absorber
material is made thinner this effective boundary layer becomes a
larger fraction of the total absorber mass and the calorimeter
response goes up. This effect was predicted by EGS3 simulation
\cite{flauger85}. It leads to  predictions for the GEM
\cite{barish92} accordion electromagnetic calorimeter (1 mm lead
and 2 mm liquid argon) that $e/mip = 0.86$ and  $e/h > 1.83$. The
Monte Carlo calculations also predict that the electromagnetic
response for liquid argon calorimeters (due to the larger $Z$
value of argon) is consistently larger than for calorimeters with
plastic-scintillator readout. The signal from neutrons ($n/mip$)
is suppressed by a  factor $0.12$ and the $n-p$ elastic
scattering products do not contribute to the signal of liquid
argon  calorimeters. These detectors only observe the $\gamma$'s
produced by inelastic neutron scattering (thermal neutron capture
escapes detection because of fast signal shaping)
\cite{wigmans87}.

To use expression  (\ref{ev7}) for reconstructing incident hadron
energies, it is necessary to know the $(e/\pi)_{Tile}$ and
$(e/\pi)_{LAr}$ ratios, which themselves depend on the hadron
energy. For this purpose,  a two cycle iteration procedure has
been developed. In the first cycle, the $(e/\pi)_{Tile}$ ratio is
iteratively evaluated using the expression:
\begin{equation}
        \Biggl(\frac{e}{\pi}\Biggr)_{Tile} =
                \frac{(e/h)_{Tile}}{1+((e/h)_{Tile}-1)
                \cdot k
           \cdot \ln{(1/e_{Tile} \cdot (e/\pi)_{Tile} \cdot R_{Tile})}}\ .
\label{epti}
\end{equation}
using the value of $(e/\pi)_{Tile}$ from a previous iteration. To
start this procedure, a value of 1.13 (corresponding to
$f_\pi^0=0.11\ln(100$ GeV$)$) has been used.

In the second cycle, the first approximation of the energy, $E$,
is calculated using the equation (\ref{ev7}) with the
$(e/\pi)_{Tile}$ ratio obtained in the first  cycle and the
$(e/\pi)_{LAr}$ ratio from equation (\ref{ev10}), where again the
iteration is initiated by  $f_{\pi^0} = 0.11 \ln(100$ GeV$)$.

\begin{figure*}[tbph]
\begin{center}
\begin{tabular}{c}
\epsfig{figure=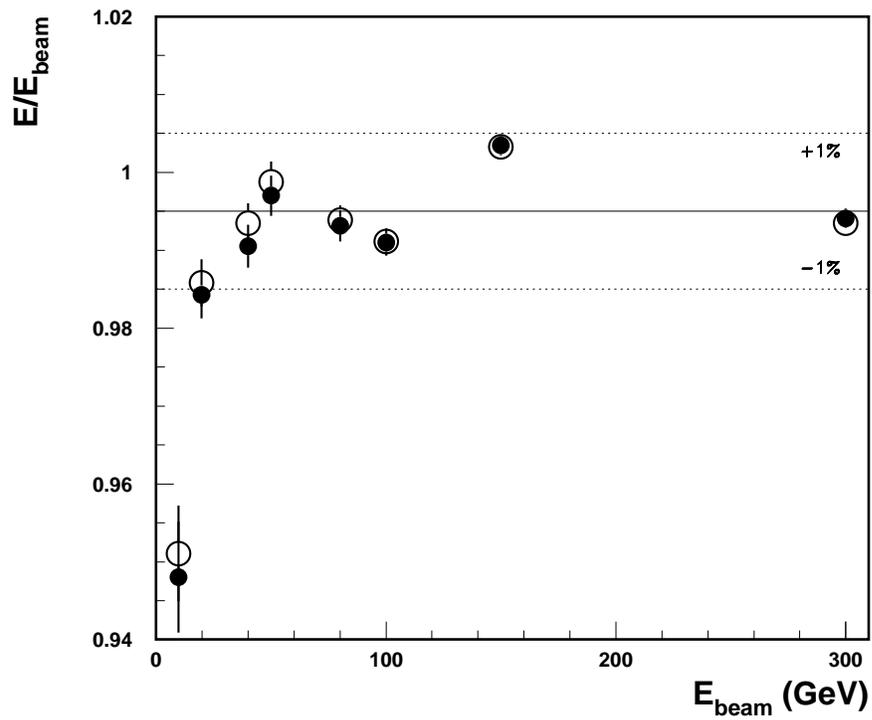,width=0.95\textwidth
}
\\
\end{tabular}
\end{center}
       \caption{
         Energy linearity as a function of the beam energy for
         the $e/h$ method  obtained using the iteration procedure
         with $\epsilon = 0.1\%$ (black circles) and with the first
         approximation (open circles).
       \label{f03-0}}
\end{figure*}

In both cycles the iterated values are arguments of a logarithmic
function; thus the iteration procedure is very fast. After the
first iteration, an accuracy of about $0.1\%$ has been achieved
for energies in the range 80$\div$150 GeV, while a second
iteration is needed to obtain the same precision for the other
beam energies. In Fig. \ref{f03-0} the energy linearity, defined
as the ratio between the mean reconstructed energy and the beam
energy,  is compared, after a first iteration,  to the linearity
obtained after iterating to a $\epsilon = 0.1\%$ accuracy,
showing a good agreement. For this reason, the suggested
algorithm of the energy reconstruction can be used for the fast
energy reconstruction in a first level trigger.

Fig.\ \ref{f03-0} also demonstrates the correctness of the mean
energy reconstruction. The mean value of $E/E_{beam}$ is equal to
$(99.5\pm0.3) \%$ and the spread is $\pm 1\%$, except for the
point at 10 GeV. However, as noted in \cite{comb96}, result at 10
GeV is strongly dependent on the effective capability to remove
events with interactions in the dead material upstream and to
separate the real pion contribution from the muon contamination.

\begin{figure*}[tbph]
\begin{center}
\begin{tabular}{cc}
\epsfig{figure=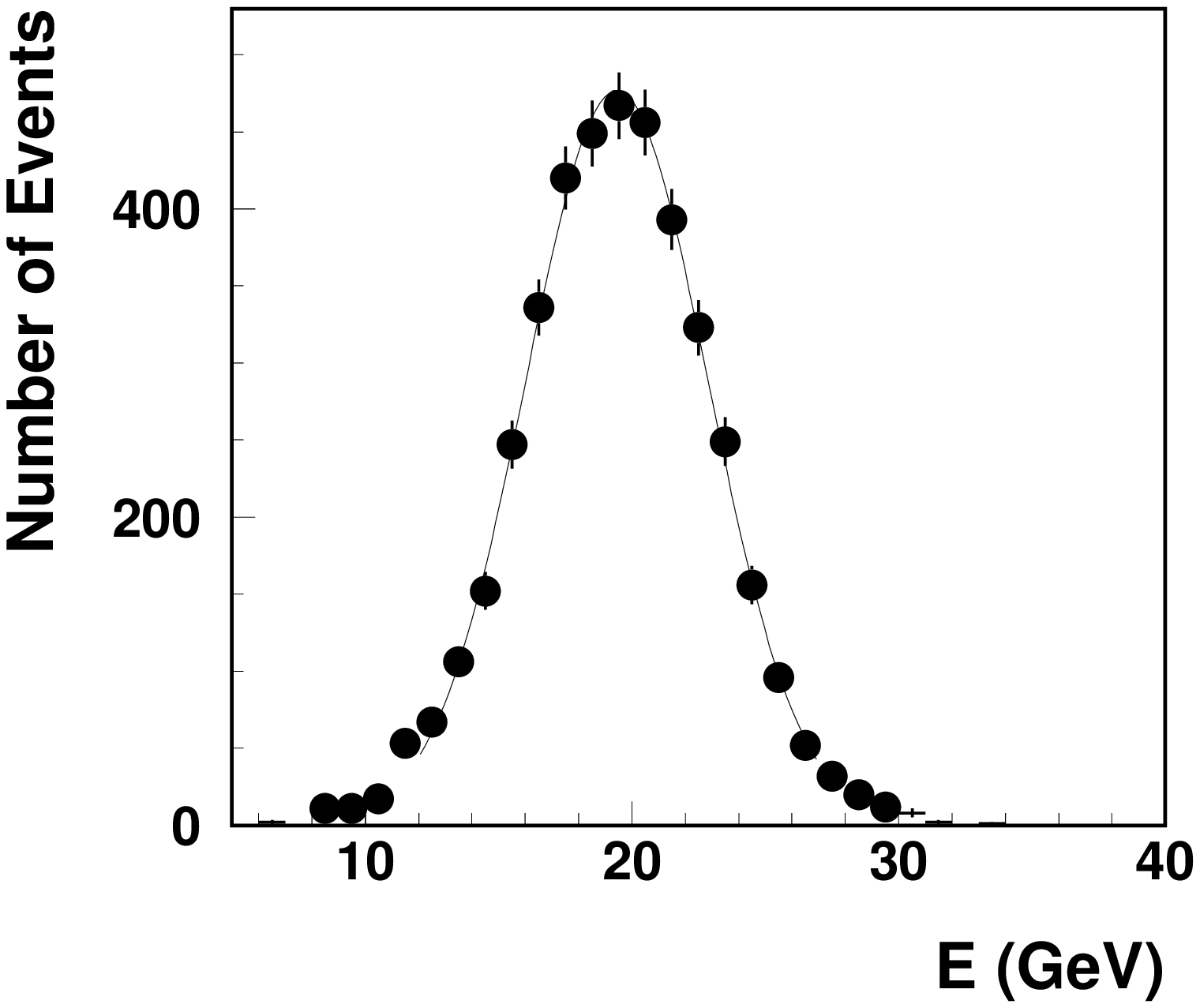,width=0.45\textwidth}
&
\epsfig{figure=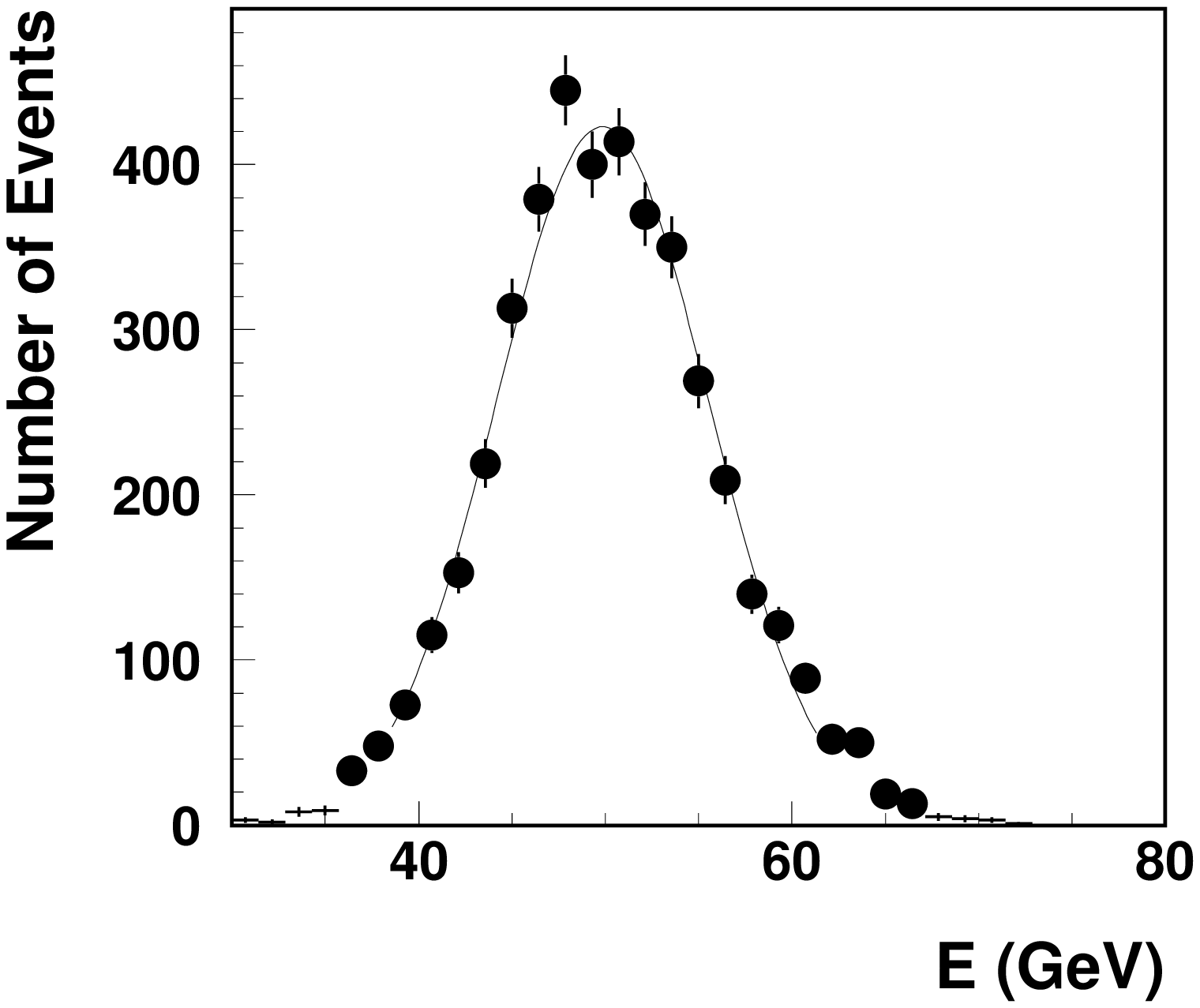,width=0.45\textwidth}
\\
\epsfig{figure=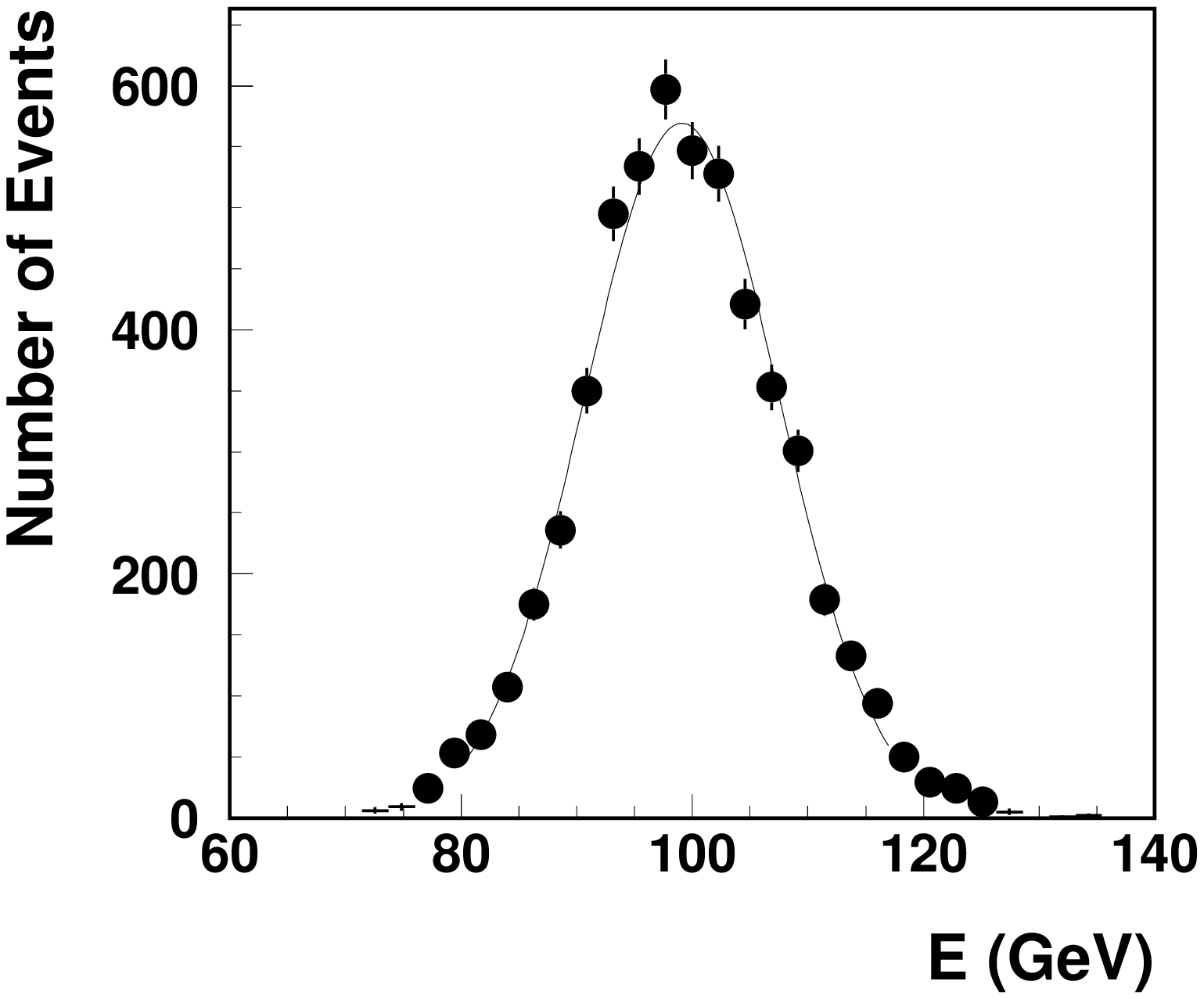,width=0.45\textwidth}
&
\epsfig{figure=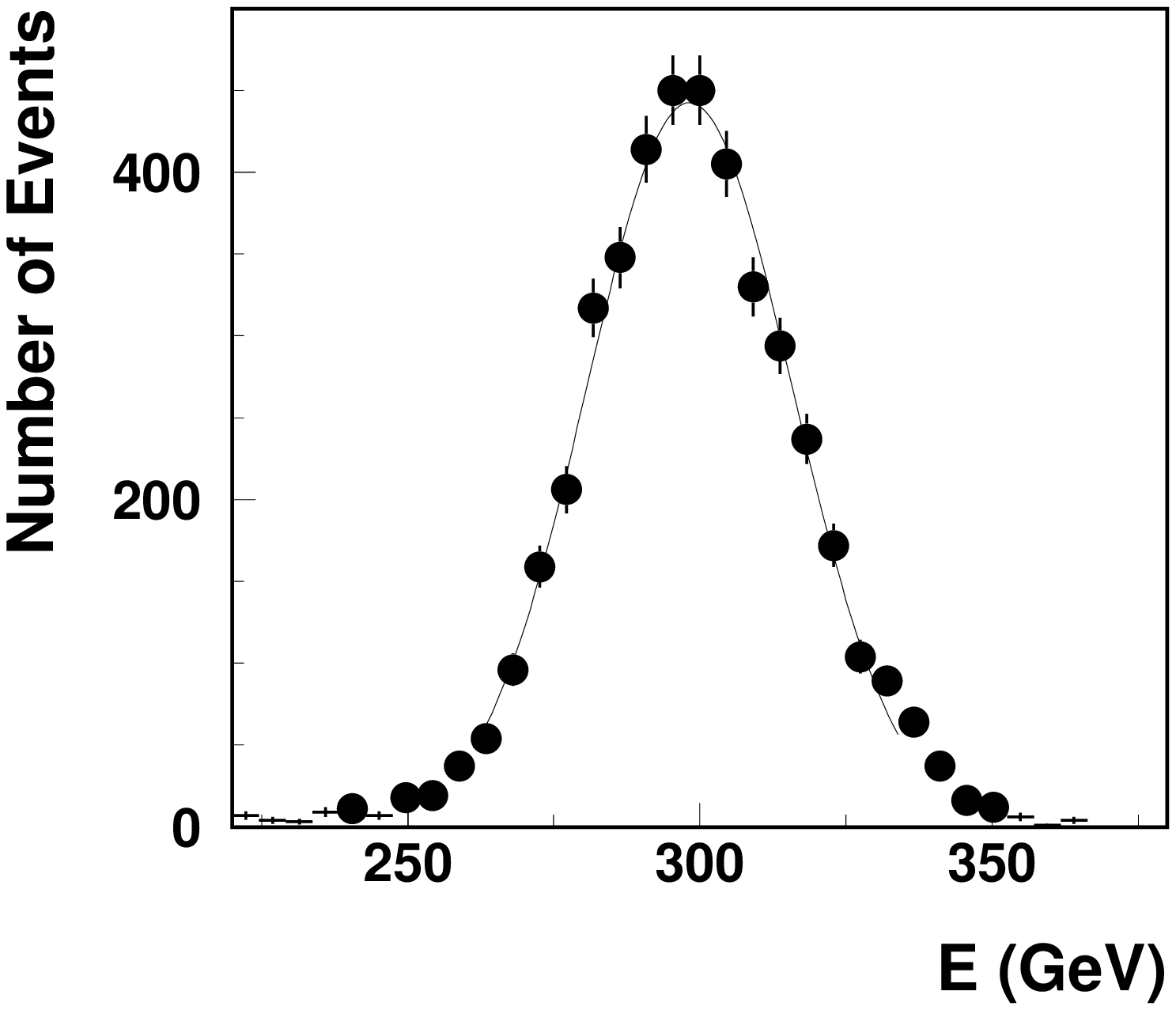,width=0.45\textwidth}
\\
\end{tabular}
\end{center}
       \caption{
                The energy distributions  for
                beam energies of 20 and 50 GeV
                (top row, left to right),
                and beam energies of 100 and 300  GeV
                (bottom row, left to right).
       \label{f01}}
\end{figure*}

Fig.\  \ref{f01} shows the pion energy spectra reconstructed with
the $e/h$ method proposed in this paper for different beam
energies. The mean and $\sigma$ values of these distributions are
extracted with Gaussian fits over $\pm 2\sigma$ range and are
reported in Table \ref{tv1} together with the fractional energy
resolution.

\begin{figure*}[tbph]
\begin{center}
\begin{tabular}{c}
\epsfig{figure=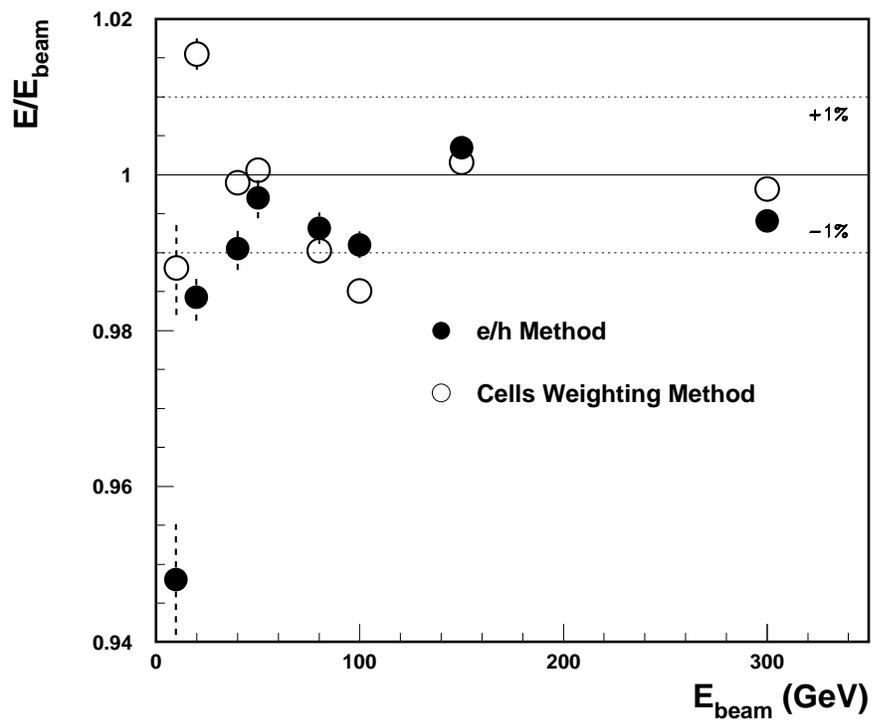,width=0.95\textwidth
}
\\
\end{tabular}
\end{center}
       \caption{
         Energy linearity as a function of the beam energy for
         the $e/h$ method
         (black circles) and the cells weighting method (open circles).
       \label{f03}}
\end{figure*}

Fig.\ \ref{f03} shows the comparison of the linearity as a
function of the beam energy for the $e/h$ method and for the
cells weighting method \cite{cavalli96}. Comparable quality of
the linearity is observed for these two methods.

\begin{figure*}[tbph]
\begin{center}
\begin{tabular}{c}
\epsfig{figure=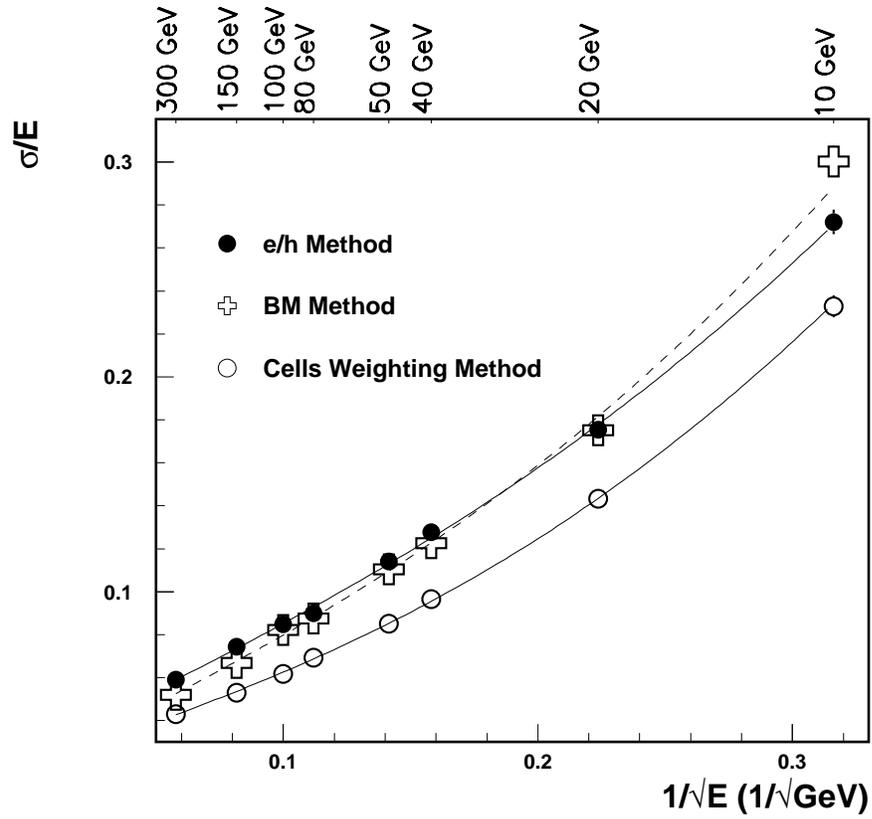,width=0.95\textwidth
}
\\
\end{tabular}
\end{center}
      \caption{
      The energy resolutions obtained with the $e/h$ method (black circles),
          the benchmark method (crosses) and the cells weighting method
         (circles).
       \label{f05}}
\end{figure*}

Fig.\ \ref{f05} shows the fractional energy resolutions ($\sigma
/E$) as a function of $1/\sqrt{E}$ obtained by three methods: the
$e/h$ method (black circles, also presented on the Table
\ref{tv1}), the ben\-chmark method \cite{comb96} (crosses), and
the cells weighting method \cite{comb96} (open circles). The
energy resolutions for the $e/h$ method are comparable with the
benchmark method and only  $30\%$ worse than for the cells
weighting method. A fit to the  data points gives the fractional
energy resolution for the $e/h$ method obtained using the
iteration procedure with $\epsilon = 0.1\%$,
\begin{eqnarray}
\sigma/E =[(58\pm3)\% /\sqrt{E}+(2.5\pm0.3)\%]\oplus (1.7\pm0.2)/E
\end{eqnarray}
for the $e/h$ method using the first approximation,
\begin{eqnarray}
\sigma/E =[(56\pm3)\% /\sqrt{E}+(2.7\pm0.3)\%]\oplus (1.8\pm0.2)/E,
\end{eqnarray}
for the benchmark method,
\begin{eqnarray}
\sigma/E =[(60\pm3)\% /\sqrt{E}+(1.8\pm0.2)\%]\oplus (2.0\pm0.1)/E,
\end{eqnarray}
and, for the cells weighting  method,
\begin{eqnarray}
\sigma/E =[(42\pm2)\% /\sqrt{E}+(1.8\pm0.1)\%]\oplus (1.8\pm0.1)/E,
\end{eqnarray}
where E is in GeV and the symbol $\oplus$ indicates a sum in
quadrature. The sampling term is consistent between the $e/h$
method and the benchmark method and is smaller by a factor of 1.5
for the cells weighting method. The constant term is the same for
the benchmark method and the cells weighting method and is larger
by $(0.7\pm0.3) \%$ for the $e/h$ method. The noise term of about
$1.8$ GeV coincide for all four cases within errors that reflect
its origin in electronic noise. Note, that from the pedestal
trigger data the total noise for the two calorimeters was
estimated to be about 1.4 GeV.

\section{Hadronic Shower Development}

\begin{figure*}[tbph]
\begin{center}
\begin{tabular}{c}
\mbox{\epsfig{figure=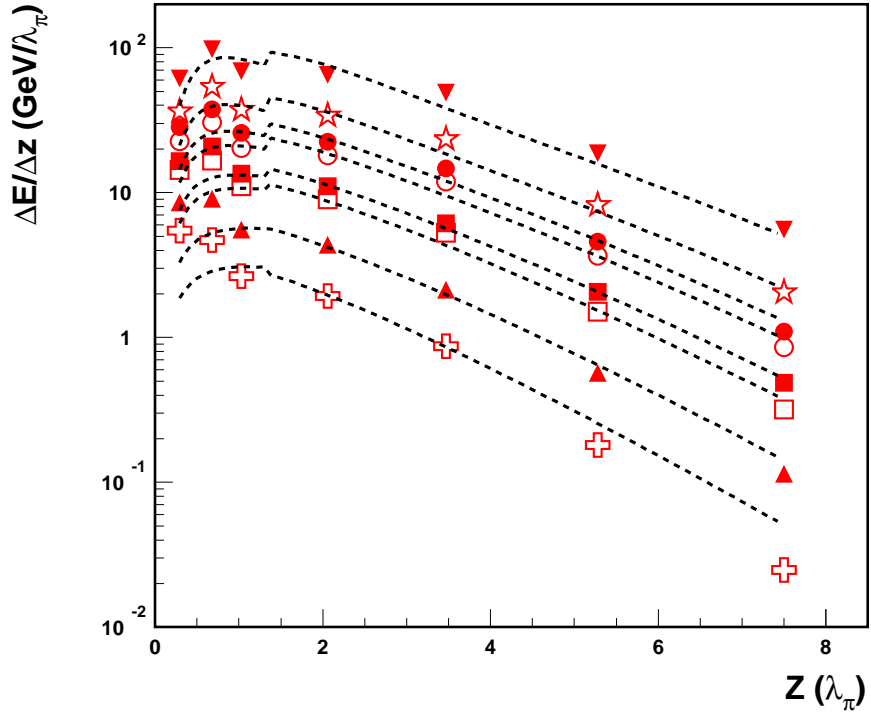,width=0.95\textwidth
}}
        \\
        \end{tabular}
\caption{The comparison between the experimental differential mean
energy depositions at 10 GeV (crosses), 20 GeV (black top
triangles), 40 GeV (open squares), 50 GeV (black squares), 80 GeV
(open circles), 100 GeV (black circles), 150 GeV (stars), 300 GeV
(black bottom triangles) and the calculated curves (\ref{elong03})
as a function of the longitudinal coordinate $z$ in units
$\lambda_{\pi}$. The errors on the data points are the errors of
the mean values and there are within symbols size.} \label{fv6-1a}
\end{center}
\end{figure*}

The $e/h$ method for energy reconstruction  has been used to study
the energy depositions, $E_i$, in each longitudinal calorimeter
sampling. Table \ref{T1} lists (and Fig.\ \ref{fv6-1a} shows) the
differential mean energy depositions $(\Delta E/ \Delta z)_i = E_i
/ \Delta z_i$ as a function of the longitudinal coordinate $z$ for
energies from 10 to 300 GeV, with $z$ expressed in interaction
length units.

A well known parameterization of the longitudinal had\-ro\-nic
sho\-wer development from the shower origin is suggested in Ref.\
\cite{bock81}:
\begin{equation}
        \frac{dE_{s} (z)}{d z} =
                N\
                \Biggl\{
                \omega\ \biggl( \frac{z}{X_0} \biggr)^{a-1}\
                e^{- b \frac{z}{X_0}}\
                + \
                (1-\omega )\
                \biggl( \frac{z}{\lambda_I} \biggr)^{a-1}\
                e^{- d \frac{z}{\lambda_I}}
                \Biggr\} \ ,
\label{elong00}
\end{equation}
where $N$ is the normalization factor, and $a,\ b,\ d,\ \omega$
are parameters ($a = 0.6165 + 0.3183\cdot \ln{E}$, $b = 0.2198$,
$d = 0.9099 - 0.0237\cdot \ln{E}$, $\omega = 0.4634$). In this
parameterization, the origin of the $z$ coordinate coincides with
shower origin, while our data are from the calorimeter face and,
due to  insufficient longitudinal segmentation, the shower origin
can not be inferred to an adequate precision. Therefore, an
analytical representation of the hadronic shower longitudinal
development from the calorimeter face has been used
\cite{kulchitsky98}:
\begin{eqnarray}
        \frac{dE (z)}{d z} & = &
                N\
                \Biggl\{
                \frac{\omega X_0}{a}
                \biggl( \frac{z}{X_0} \biggr)^a
                e^{- b \frac{z}{X_0}}
                {}_1F_1 \biggl(1,a+1,
                \biggl(b - \frac{X_0}{\lambda_I} \biggr) \frac{z}{X_0}
                \biggr)
                \nonumber \\
                & & + \
                \frac{(1 - \omega ) \lambda_I}{a}
                \biggl( \frac{z}{\lambda_I} \biggr)^a
                e^{- d \frac{z}{\lambda_I}}
                {}_1F_1 \biggl(1,a+1,
                \bigl( d -1 \bigr) \frac{z}{\lambda_I} \biggr)
                \Biggr\} ,
\label{elong03}
\end{eqnarray}
where ${}_1F_1(\alpha, \beta, z)$ is the confluent hypergeometric
function. Note that the formula (\ref{elong03}) is given for a
calorimeter characterized by its $X_0$ and $\lambda_I$. In the
combined setup, the values of $X_0$, $\lambda_I$ and the $e/h$
ratios are different for electromagnetic and hadronic
compartments. So, the use of formula (\ref{elong03}) is not
straightforward for the description of the hadronic shower
longitudinal profiles.

To overcome this problem, Ref.\ \cite{kulchitsky99} suggests an
algorithm to combine the electromagnetic ca\-lo\-ri\-me\-ter
($em$) and hadronic calorimeter ($had$) curves of the
differential longitudinal energy deposition $dE/dz$. At first,
the mean hadronic shower develops according eq.\ (\ref{elong03})
in the electromagnetic calorimeter to the boundary value $z_{em}$
which corresponds to a certain integrated measured energy
$E_{em}(z_{em})$. Then, using the corresponding integrated
hadronic curve, $E(z)=\int_0^z (dE/dz) dz$, the point $z_{had}$
is found from the equation $E_{had}(z_{had}) = E_{em}(z_{em})+
E_{dm}$. From this point a shower continues to develop in the
hadronic calorimeter. In principle, instead of the measured value
of $E_{em}$ one can use the calculated value of $E_{em} =
\int_0^{z_{em}} (dE/dz) dz$ obtained from the integrated
electromagnetic curve. The combined curves have been obtained in
this manner.

Fig.\ \ref{fv6-1a} shows the differential energy depositions
$(\Delta E/ \Delta z)_i = E_i / \Delta z_i$ as a function of the
longitudinal coordinate $z$ in units of $\lambda_{\pi}$ for the
energy from 10 to 300 GeV and a comparison with the combined
curves for the longitudinal hadronic shower profiles (dashed
lines). The level of agreement was estimated using the function
$\chi^2$ where, following Ref.\ \cite{bock81},  the variances of
the energy depositions are taken to be equal to the depositions
themselves. A significant disagreement ($P(\chi^2) < 0.1\%$) has
been observed between the experimental data and the combined
curves in the region of the LAr calorimeter, especially at low
energies.

We attempted to improve the description and to include such
essential feature of a ca\-lo\-ri\-me\-ter as the $e/h$ ratio.
Several modifications and adjustments of some parameters of the
parameterization (\ref{elong03}) have been tried. The conclusion
is that replacing the two parameters $b$ and $\omega$ in the
formula (\ref{elong03}) with $b = 0.22 \cdot (e/h)_{cal} /
(e/h)_{cal}^{\prime}$ and $\omega = 0.6  \cdot (e/\pi)_{cal} /
(e/\pi)_{cal}^{\prime}$ results in a  reasonable description of
the experimental data. Here the values of the
$(e/h)_{cal}^{\prime}$ ratios are $(e/h)^{\prime}_{em} \approx
1.1$ and $(e/h)^{\prime}_{had} \approx 1.3$ which correspond to
the data used for the Bock et al.\ parameterization \cite{bock81}.
The $(e/\pi)_{cal}^{\prime}$ are calculated using formulas
(\ref{ev10}) and (\ref{ev10-0}).

\begin{figure*}[tbph]
\begin{center}
\begin{tabular}{c}
\mbox{\epsfig{figure=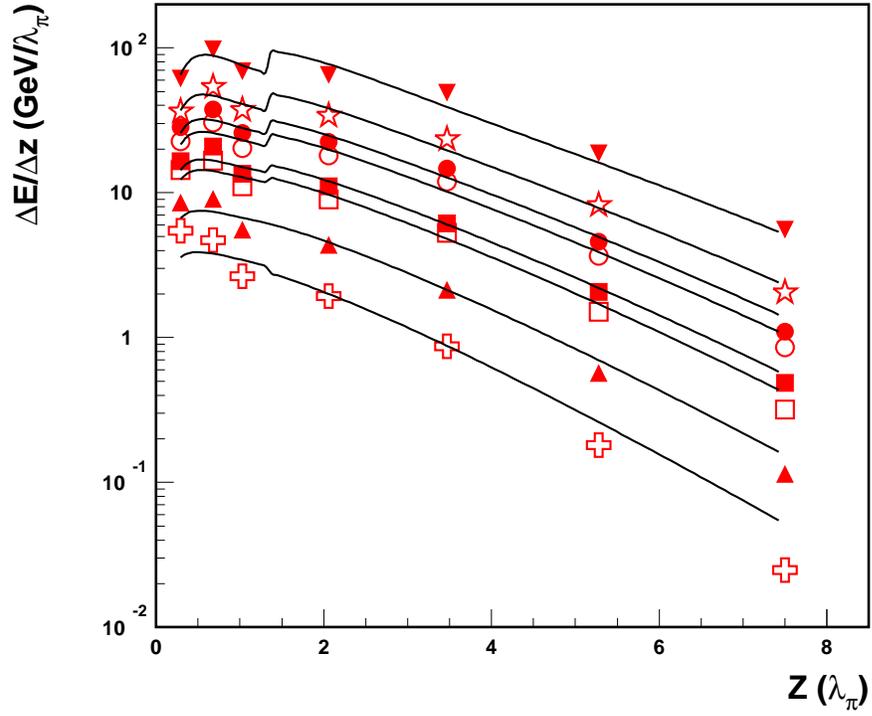,width=0.95\textwidth
}}
        \\
        \end{tabular}
\caption{The experimental differential mean longitudinal energy
depositions at 10 GeV (crosses), 20 GeV (black top triangles), 40
GeV (open squares), 50 GeV (black squares), 80 GeV (open
circles), 100 GeV (black circles), 150 GeV (stars), 300 GeV
(black bottom triangles) energies as a function of the
longitudinal coordinate $z$ in units $\lambda_{\pi}$ for the
combined calorimeter and the results of the description by the
modified parameterization for the equation (\ref{elong03}).}
\label{fv6-1b}
\end{center}
\end{figure*}

In Fig.\ \ref{fv6-1b} the experimental differential longitudinal
energy depositions and the results of the description by the
modified parameterization (solid lines) are compared. There is a
reasonable agreement (the probability of description is more than
$5\%$) between the experimental data and the curves. Note, that
previous comparisons between Monte-Carlo and data have shown that
FLUKA describes well the longitudinal shape of hadronic showers
\cite{comb96}.

\begin{figure*}[tbph]
\begin{center}
\begin{tabular}{c}
\epsfig{figure=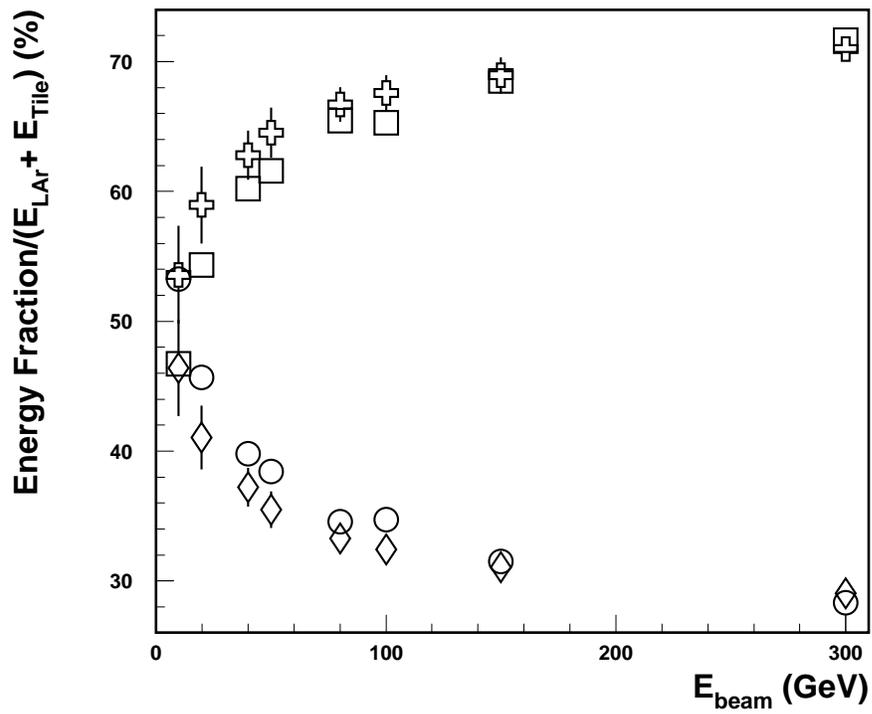,width=0.95\textwidth
}
\\
\end{tabular}
\end{center}
\caption{ Energy deposition (percentage) in the LAr and Tile
calorimeters at different beam energies. The circles (squares)
are the measured energy depositions in the LAr (Tile)
calorimeter, the diamonds (crosses) are the calculated energy
depositions. } \label{f04-a}
\end{figure*}

The obtained parameterization has some additional applications.
For example, this formula may be used for an estimate of the
energy deposition in various parts of a combined calorimeter.
This is demonstrated in Fig.\ \ref{f04-a} in which the measured
and calculated relative values of the energy deposition in the
LAr and Tile calorimeters are presented. The errors of the
calculated values presented in this figure reflect the
uncertainties of the parameterization (\ref{elong03}). The
relative energy deposition in the LAr calorimeter decreases from
about 50\% at 10 GeV to 30\% at 300 GeV. Conversely, the fraction
in the Tile calorimeter increases as the energy increases.

\section{Conclusions}

Hadron energy reconstruction for the ATLAS barrel prototype
combined ca\-lo\-ri\-me\-ter has been carried out in the
framework of the non-pa\-ra\-met\-ri\-cal method. The
non-parametrical method of the energy reconstruction for a
combined ca\-lo\-ri\-me\-ter uses only the $e/h$ ratios and the
electron calibration constants, without requiring the
determination of other parameters by a minimization technique.
Thus, it can be used for the fast energy reconstruction in a
first level trigger. The value of the $e/h$ ratio obtained  for
the electromagnetic compartment of the combined
ca\-lo\-ri\-me\-ter is $1.74\pm0.04$ and agrees with the
prediction that $e/h > 1.66$ for this calorimeter. The ability to
reconstruct the mean values of particle energies (for energies
larger than 10 GeV) within $\pm 1\%$ has been demonstrated. The
obtained fractional energy resolution is $[(58\pm3)\%
/\sqrt{E}+(2.5\pm0.3)\%]\oplus (1.7\pm0.2)/E$. The results of the
study of the longitudinal hadronic shower development have also
been presented.

\section{Acknowledgments}

We would like to thank the technical staffs of the collaborating
Institutes for their important and timely contributions.
Financial support is acknowledged from the funding agencies of the
collaborating Institutes. Finally, we are grateful to the staff
of the SPS, and in particular to Konrad Elsener, for the
excellent beam conditions and assistance provided during our
tests.




\newpage
\begin{table}[tbhp]
\begin{center}
\caption{ Mean reconstructed energy, energy resolution and
fractional energy resolution for the various beam energies.}
\label{tv1}
\begin{tabular}{|r|c|c|c|}
\multicolumn{4}{c}{\mbox{~~}}\\[-3mm]
\hline
$E_{beam}$&$E$ (GeV)&$\sigma$ (GeV)&$\sigma / E\ (\%)$\\
\hline
$10^{\ast}$     GeV&  $9.30\pm0.07$& $2.53\pm0.05$&$27.20\pm0.58$\\
$20^{\star}$    GeV& $19.44\pm0.06$& $3.41\pm0.06$&$17.54\pm0.31$\\
40              GeV& $39.62\pm0.11$& $5.06\pm0.08$&$12.77\pm0.21$\\
50              GeV& $49.85\pm0.13$& $5.69\pm0.13$&$11.41\pm0.26$\\
80              GeV& $79.45\pm0.16$& $7.14\pm0.14$& $8.99\pm0.18$\\
100             GeV& $99.10\pm0.17$& $8.40\pm0.16$& $8.48\pm0.16$\\
150             GeV&$150.52\pm0.19$&$11.20\pm0.18$& $7.44\pm0.12$\\
300             GeV&$298.23\pm0.37$&$17.59\pm0.33$& $5.90\pm0.11$\\
\hline
\multicolumn{4}{@{}l@{}}{$^{\ast}$The measured value of the
                                               beam energy is 9.81 GeV.}\\
\multicolumn{4}{@{}l@{}}{$^{\star}$The measured value of the
                                        beam energy is 19.8 GeV.}     \\
\end{tabular}
\end{center}
\end{table}

\begin{table}[tbhp]
\vspace*{-10mm}
\begin{center}
\caption{ The differential mean energy depositions $\Delta E/
\Delta z$ ($GeV/\lambda_\pi$) as a function of the longitudinal
coordinate $z$ for the various beam energies.} \label{T1}
\begin{tabular}{|c|c|c|c|c|c|}
\multicolumn{6}{c}{\mbox{~~}}\\[-3mm]
\hline
    N&  z&\multicolumn{4}{|c|}{$E_{beam}$ (GeV)}\\
\cline{3-6}
depth&($\lambda_{\pi}$)&10  &20           &40       &50                 \\
\hline
1  &0.294&$5.45\pm0.08$  & $8.58\pm0.16$ &$14.3\pm0.2$   &$16.6\pm0.4$ \\
2  &0.681&$4.70\pm0.08$  & $9.10\pm0.15$ &$16.7\pm0.2$   &$20.8\pm0.3$ \\
3  &1.026&$2.66\pm0.06$  & $5.55\pm0.11$ &$11.1\pm0.2$   &$13.6\pm0.2$ \\
4  &2.06 &$1.93\pm0.03$  & $4.35\pm0.06$ &$8.99\pm0.08$  &$11.0\pm0.1$ \\
5  &3.47 &$0.87\pm0.02$  & $2.13\pm0.04$ &$5.29\pm0.06$  &$6.15\pm0.10$ \\
6  &5.28 &$0.18\pm0.01$  & $0.57\pm0.02$ &$1.50\pm0.03$  &$2.07\pm0.05$ \\
7  &7.50 &$0.025\pm0.003$& $0.11\pm0.01$ &$0.32\pm0.01$  &$0.49\pm0.02$\\
\hline
    N&  z&\multicolumn{4}{|c|@{}}{$E_{beam}$ (GeV)}\\
\cline{3-6}
depth&($\lambda_{\pi}$)&80&100&150&300 \\
\hline
1  &0.294&$22.6\pm0.6$ &$28.4\pm0.6$ &$36.3\pm0.7$ &$61.3\pm1.5$        \\
2  &0.681&$30.4\pm0.4$ &$37.6\pm0.5$ &$53.5\pm0.8$ &$97.9\pm1.7$        \\
3  &1.026&$20.3\pm0.3$ &$25.7\pm0.4$ &$37.2\pm0.6$ &$68.9\pm1.2$        \\
4  &2.06 &$18.0\pm0.1$ &$22.4\pm0.2$ &$33.9\pm0.3$ &$64.8\pm0.7$        \\
5  &3.47 &$11.9\pm0.1$ &$14.6\pm0.2$ &$23.3\pm0.2$ &$49.0\pm0.5$        \\
6  &5.28 &$3.66\pm0.06$&$4.57\pm0.08$&$8.18\pm0.13$&$18.6\pm0.3$        \\
7  &7.50 &$0.86\pm0.03$&$1.10\pm0.04$&$2.04\pm0.06$&$5.54\pm0.15$       \\
\hline
\end{tabular}
\end{center}
\end{table}
\clearpage

\clearpage

\end{document}